\documentclass[12pt]{article}
\newcommand{\FlipTR}{UCR-TR-2022-FLIP-NCC-2000}
\usepackage{amsmath}
\usepackage{amssymb}
\usepackage{amsfonts}
\usepackage{graphicx}
\usepackage[utf8]{inputenc}     % inspire bibs
\usepackage{aas_macros}				  % ads bibs
\usepackage{bm}                 % \boldsymbol
\usepackage{amsthm}
\usepackage{microtype}

\usepackage{slashed}				% \slashed{k}
\usepackage{mathrsfs}				% Weinberg-esque letters
\usepackage{bbm}					% \mathbbm{1} conflict: XeLaTeX 
\usepackage{cancel}					
\usepackage[dvipsnames]{xcolor}
\usepackage[hang,flushmargin]{footmisc} % no footnote indent

\usepackage{fancyhdr}		% preprint number
\usepackage{lipsum}			% block of text 
\usepackage{framed}			% boxed remarks
\usepackage{subcaption}	% subfigures
\usepackage{cite}			  % group cites
\usepackage{xspace}			% macro spacing

\usepackage{booktabs}		% professional tables
\usepackage{multirow}		% multirow elements in a table
\usepackage[font=small]{caption} 	% caption font is small
\usepackage{float}         			  % for strict placement e.g. [H]

\usepackage[margin=2.5cm]{geometry} % margins
\graphicspath{{figures/}}			      % figure folder
\numberwithin{equation}{section}    % set equation numbering
\usepackage{multicol}
\usepackage{etoolbox}
\usepackage{relsize}
\patchcmd{\thebibliography}
  {\list}
  {\begin{multicols}{2}\smaller\list}
  {}
  {}
\appto{\endthebibliography}{\end{multicols}}

\let\oldenumerate\enumerate
\renewcommand{\enumerate}{
  \oldenumerate
  \setlength{\itemsep}{4pt}
  \setlength{\parskip}{0pt}
  \setlength{\parsep}{0pt}
}

\let\olditemize\itemize
\renewcommand{\itemize}{
  \olditemize
  \setlength{\itemsep}{4pt}
  \setlength{\parskip}{0pt}
  \setlength{\parsep}{0pt}
}

\usepackage{lineno}
\usepackage{scalefnt} 
\newcommand\acro[1]{{\small {#1}}}

\renewcommand{\tilde}{\widetilde}   % tilde over characters
\renewcommand{\text}{\textnormal}	% text in equations 
\newcommand{\email}[1]{\href{mailto:#1}{#1}}
\newenvironment{institutions}[1][2em]{\begin{list}{}{\setlength\leftmargin{#1}\setlength\rightmargin{#1}}\item[]}{\end{list}}

\fancypagestyle{firststyle}{
	\rhead{\footnotesize%
	\texttt{\FlipTR}%
	}}

\usepackage{etoc}

  \newcommand{\massmixing}{\kappa} % "sine-beta-squared-bar"
   
  \usepackage{tabularx}

\newcommand{\Uone}{\acro{U(1)}\xspace}

\usepackage[
	colorlinks=true,
	citecolor=green!50!black,
	linkcolor=NavyBlue!75!black,
	urlcolor=green!50!black,
	hypertexnames=false]{hyperref}

\usepackage{orcidlink}

\def\beq{\begin{equation}}
\def\eeq#1{\label{#1}\end{equation}}
\def\eeqn{\end{equation}}
\def\beqa{\begin{eqnarray}}
\def\eeqa#1{\label{#1}\end{eqnarray}}
\def\eeqan{\end{eqnarray}}

\def\leqn#1{(\ref{#1})}

\usepackage{physics} 

\begin{document}

\thispagestyle{firststyle}

\begin{center}

    {\LARGE \textbf{Dark\,$Z$ at the International Linear Collider}\par}
    \vspace{.3em}
    \vskip .5cm
   { \bf 
   	Yik Chuen San$^{a}$,
   	Maxim Perelstein$^{a}$,
   	and
   	Philip Tanedo$^{b}$  
   	} 
   \\ 
   \vspace{-.2em}
   { \tt \footnotesize
	    \email{ys828@cornell.edu},
	    \email{m.perelstein@cornell.edu}\orcidlink{0000-0002-9965-3911},
	    \email{flip.tanedo@ucr.edu}\orcidlink{0000-0003-4642-2199}
   }
	
   \vspace{-.2cm}

   \begin{institutions}[1.5cm]
   \footnotesize
   $^{a}$ 
   {\it 
       Laboratory for Elementary Particle Physics, 
       Cornell University, Ithaca, \acro{NY} 14853%, \acro{USA}
       }
	\\ 
	\vspace*{0.05cm}   
	$^{b}$ 
	{\it 
	    Department of Physics \& Astronomy, 
	    University of  California, Riverside, 
	    \acro{CA} 92521	    
	    }  
   \end{institutions}

\end{center}

%%%%%%%%%%%%%%%%%%%%%
%%%  ABSTRACT    %%%%
%%%%%%%%%%%%%%%%%%%%%

\begin{abstract}
\noindent 
A dark\,$Z$ is a massive Abelian gauge boson which is coupled to the Standard Model through both kinetic and mass mixing with the electroweak sector. We study the phenomenology of the dark\,$Z$ at an energy-frontier $e^+e^-$ collider, such as the proposed International Linear Collider (\acro{ILC}). We show that precision electroweak constraints and the current bounds from the hadron colliders allow a dark\,$Z$ that is kinematically accessible at the \acro{ILC} with 250~GeV or 500~GeV center-of-mass energy. Further, the reach of the \acro{ILC} searches for a dark\,$Z$ significantly exceeds the expected reach of the high luminosity \acro{LHC}. If a signal consistent with a dark\,$Z$ is discovered, it would motivate a dedicated run of the \acro{ILC} at the center-of-mass energy matching the dark\,$Z$ mass. We demonstrate that a short one month run at design luminosity could measure the dark\,$Z$ chiral couplings to fermions with percent precision. This measurement can be used to discriminate between competing theoretical models of the resonance: for example, the dark $Z$ can be distinguished from a dark photon with purely kinetic mixing.  
\end{abstract}

\small
\setcounter{tocdepth}{2}
\tableofcontents
\normalsize
%\clearpage

%%%%%%%%%%%%%%%%%%%%%
%%%  THE CONTENT  %%%
%%%%%%%%%%%%%%%%%%%%%
\newpage

\section{Introduction} 

The discovery of the Higgs boson at the Large Hadron Collider (\acro{LHC}) strongly motivates the construction of a Higgs factory: an electron-positron collider with center-of-mass energy in the 250--500~GeV range. Several proposals are currently under consideration~\cite{Baer:2013cma,Fujii:2017vwa,Linssen:2012hp,CLICdp:2018cto,CEPCStudyGroup:2018ghi,FCC:2018byv,FCC:2018evy}, and hopes are high that at least one of them will be realized in the next one to two decades. While the main motivation for these colliders is a comprehensive and precise study of the Higgs, they can also search for physics Beyond the Standard Model in ways complementary to the \acro{LHC}. It is this aspect of the physics program that we focus on in this paper.  

Additional gauge interactions are ubiquitous in well-motivated extensions of the Standard Model (\acro{SM}), such as grand unified theories and string theory. A particularly simple example is an extra \Uone gauge group. When \acro{SM} fermions are charged under the extra \Uone, the collider phenomenology of the corresponding gauge boson, the $Z^\prime$, is well-studied. Unfortunately, the \acro{LHC} sets a lower limit on the $Z^\prime$ mass in the multi-TeV range, precluding the possibility of its direct production at next-generation $e^+e^-$ colliders; although the $Z^\prime$ may still be detected through its indirect effects. We focus on an equally well-motivated scenario that has received less attention so far: the \acro{SM} fermions are \emph{not} directly charged under the extra \Uone. The new gauge boson, called a dark $Z$, can still interact with the \acro{SM} through kinetic mixing with hypercharge~\cite{Kobzarev:1966qya,Okun:1982xi,Holdom:1985ag,Holdom:1986eq}, as well as mass mixing induced by an extended Higgs sector~\cite{Foot:1991kb, Davoudiasl:2012ag}. (If only hypercharge mixing is present, this boson is known as a dark photon.) Such interactions are not constrained by charge quantization, and can be significantly weaker than \acro{SM} gauge couplings. We show that the current constraints from the \acro{LHC} and other experiments are consistent with models in which the dark $Z$ is accessible to a direct search at a realistic 250--500~GeV $e^+e^-$ collider. We use the International Linear Collider (\acro{ILC}) design parameters as the collider benchmark in our study. Further, we extrapolate the current \acro{LHC} sensitivity to the 3~ab$^{-1}$ data set expected at the high-luminosity (\acro{HL})-\acro{LHC} to show that there is a large region of parameter space where the $e^+e^-$ collider can make the first observation of the dark $Z$.   

If a particle consistent with a dark\,$Z$ is discovered, the next step would be to measure its properties, such as mass and couplings to the \acro{SM}. An $e^+e^-$ collider can perform these measurements with unparalleled precision. Unlike the \acro{LHC}, a lepton collider can separately determine the dark\,$Z$ couplings to left-handed and right-handed \acro{SM} fermions. These chiral couplings are crucial for discriminating between different possible models of the new particle.
We use a benchmark dark-$Z$ model as an illustrative example to show that a relatively short (one or two months) dedicated run where the $e^+e^-$ collision center-of-mass energy is matched to the dark\,$Z$ mass can measure the couplings with percent-level errors, unambiguously establish the presence of parity violation in the dark-$Z$ couplings, and probe the underlying sources of dark-$Z$ interactions with the \acro{SM}. We show that the $e^+e^-$ collider may distinguish between pure kinetic mixing versus models with mass mixing through exotic Higgs fields. The program of on-resonance measurements of the dark $Z$ that we outline is a ``dark'' counterpart to the precision electroweak program in the 1990s at \acro{LEP} and \acro{SLC}. However, there are important differences that need to be properly taken into account, such as the very small intrinsic width of the dark\,$Z$ compared to the \acro{SM} $Z$.          
 
Before proceeding, let us make the following observation. While dark gauge bosons have attracted much attention in recent years, a vast majority of this work has been focused on mass scales in the MeV--GeV range~\cite{Alexander:2016aln,Battaglieri:2017aum,Fabbrichesi:2020wbt}. In contrast, we  investigate much heavier dark gauge bosons with masses around the weak scale. From a theoretical point of view, this mass is a free parameter and any scale is equally well motivated. Part of the motivation for sub-GeV dark photons comes from the observation that they can serve as a mediator between the \acro{SM} and a light dark matter particle, and produce a thermal relic density of dark matter consistent with observations. The dark\,$Z$ studied here cannot play this role in a minimal model. However, slightly more complicated models of the dark sector ({\it e.g.} involving a state directly coupled to the dark\,$Z$ which decays to the stable dark matter state as in the super-\acro{WIMP} mechanism~\cite{Feng:2003xh}) can easily accommodate the observed dark matter with a weak-scale dark $Z$ mediator. Thus, a comprehensive exploration of the dark sector has to include the possibility of weak-scale dark gauge bosons, and energy-frontier colliders offer the only way to access this regime experimentally.   

The rest of the paper is organized as follows. Section~\ref{sec:simplified:model} describes the theoretical model underlying our study. In Section~\ref{sec:exper} we discuss the current constraints on the model from precision electroweak fits and direct searches at the \acro{LHC}, as well as projected reach of future searches at the \acro{HL-LHC} and the \acro{ILC}. The analysis of this section extends previous studies of dark photon phenomenology at TeV-scale colliders~\cite{Curtin:2014cca,Karliner:2015tga} to include the effects of mass mixing in the dark\,$Z$ model. In Section~\ref{sec:Zpole:Obs}, we explore the physics potential of a short, dedicated run of the \acro{ILC} with the center-of-mass energy at the dark\,$Z$ resonance. We conclude in Section~\ref{sec:conc}, while the Appendix contains some of the details of our model and its ultraviolet completion, as well as explicit formulas for the ``precision dark\,$Z$" observables.

\section{\texorpdfstring{Three-Parameter Dark\,$Z$ Model}{Three-Parameter Dark Z Model}}
\label{sec:simplified:model}

We present a model of a new spin-1 particle with both kinetic and mass mixing. The parameters are the dark\,$Z$ mass $m_{A'}$, the kinetic mixing $\varepsilon$, and a dimensionless mass mixing parameter $\massmixing$. The theory is a limit of a type-1 two-Higgs doublet model augmented with a third electroweak-neutral dark Higgs.
Refs.~\cite{Gopalakrishna:2008dv, Davoudiasl:2012ag} present an alternative two-Higgs doublet model that realizes the same low-energy theory. While our model is less minimal, it cleanly separates the dimensionful parameters that control the dark\,$Z$ mass and mass mixing. We provide additional details of the effective theory in Appendix~\ref{app:model:summaru}.

\subsection{\texorpdfstring{Kinetic Mixing: $\varepsilon$}{Kinetic Mixing}}

We assume an Abelian `dark' gauge symmetry \acro{U(1)$_\text{d}$} with gauge boson $A'_\mu$. Standard Model particles not uncharged under this symmetry. The gauge kinetic terms contain a mixing term with the hypercharge boson $B_\mu$, 
\begin{align}
  \mathcal L_\text{gauge}
  &=
  - \frac{1}{4} B_{\mu \nu} B^{\mu \nu}
  + \frac{\varepsilon}{2 c_W} B_{\mu \nu} A'^{\mu \nu}
  - \frac{1}{4} A'_{\mu \nu} A'^{\mu \nu} \ .
  \label{eq:kinetic:mixing}
\end{align}
This mixing may be generated by loops of heavy particles charged under both Abelian groups~\cite{Kobzarev:1966qya,Okun:1982xi,Holdom:1985ag,Holdom:1986eq}.
We normalize $\varepsilon$ by the cosine of the Weinberg angle
$c_W=\cos\theta_W$ 
so that the kinetic mixing with the electromagnetic field strength is $\varepsilon$. In what follows, we adopt the notation $s_W = \sin\theta_W$ and $t_W = \tan\theta_W$.
We diagonalize the gauge kinetic term by transforming
\begin{align}
B &\to B + \frac{\varepsilon}{c_W}\mathcal N A'
&
A' &\to \mathcal N A' \ .
\label{eq:diagonalization:1:kinetic:mixing}
\end{align}
Henceforth we set the normalization factor $\mathcal N^{-2} = 1-\varepsilon^2/c_W^2 \approx 1$ because the correction is of higher order in $\varepsilon$ compared to the accuracy required for this study.

\subsection{\texorpdfstring{Symmetry Breaking and Mass--Mixing: $\kappa$}{Symmetry Breaking and Mass--Mixing}}
\label{sec:symmetry:breaking:extended:higgs}

The \acro{U(1)$_\text{d}$} and electroweak symmetry are broken by separate vacuum expectation values (vevs), $v_\text{d}$ and $v_\text{EW}$, as well as by a mixed vev, $v_\text{mix} \ll v_\text{EW}$. A benchmark \acro{UV} model carries an extended Higgs sector with the following quantum numbers:
\begin{align}
  H_\text{EW} &= \square_{\frac{1}{2},0} 
  &
  % H_m &= \square_{\frac{1}{2},1} 
  H_\text{mix} &= \square_{\frac{1}{2},q_\text{d}} 
  &
  H_\text{d} &= \mathbbm{1}_{0,-q_\text{d}}  \ ,
  \label{eq:Higgses:quantum:numbers}
\end{align}
where $\square$ or $\mathbbm{1}$ refer to an \acro{SU(2)$_\text{L}$} doublet or singlet and the subscripts refer to the \acro{U(1)$_Y$} hypercharge and \acro{U(1)$_\text{d}$} dark charges.
We assume that the extended Higgs sector potential has large enough quartic couplings relative to the dark gauge coupling that the non-Goldstone modes decouple from the dark\,$Z$ dynamics; see Appendix~\ref{app:UV:model}. (Ref.~\cite{Liu:2017lpo} studied the phenomenology of a model with a non-decoupled scalar.)

The vevs of the Higgs fields give produce gauge boson mass terms,
\begin{align}
  \mathcal L%_\text{mass}
  \supset &
  \left[
    \left(
      \frac{1}{2} g' B 
      - \frac{1}{2} g W^3
    \right)
    \frac{v_\text{EW}}{\sqrt{2}}
  \right]^2
  +
  \left[
    \left(
      \frac{1}{2} g' B 
      - \frac{1}{2} g W^3 
      + g_\text{d} q_\text{d} A'
    \right) 
    \frac{v_\text{mix}}{\sqrt{2}}
  \right]^2
  +
  \left[
    g_\text{d}q_\text{d}  A' \frac{v_\text{d}}{\sqrt{2}}
  \right]^2 
\ .
  \label{eq:mass:terms:from:vevs}
\end{align}
The $v_\text{mix}$ term is a mass mixing between the dark\,$Z$ and Standard Model~$Z$. The net order parameter of electroweak symmetry breaking is $v^2 = v_\text{EW}^2 + v_{\rm mix}^2 =(246~\text{GeV})^2$. 
The eigenvalues of the mass-squared matrix are the dark and Standard Model $Z$ squared masses, $m_{A'}^2$ and $m_Z^2 = (91~\text{GeV})^2$. The following parameterization of mass mixing is particularly convenient,
\begin{align}
   \massmixing &= 2q_\text{d} \frac{g_\text{d}}{\sqrt{g'^2+g^2}}\frac{v_\text{mix}^2}{v^2} 
   &
   \text{mass mixing parameter}
  \label{eq:mass:mixing:definition}
\end{align}
We will assume that $\kappa\ll 1$, as is in fact required by precision electroweak observables, discussed below. Moreover, we treat $\kappa$ and $\varepsilon$ as small parameters of the same order. Defining $c_\delta = \cos\delta$ and $s_\delta = \sin\delta$, the transformation to the mass eigenstates is then
\begin{align}
  \begin{pmatrix}
    A'\\Z
  \end{pmatrix}
  &=
  \begin{pmatrix}
    \phantom{+}c_\delta & s_\delta \\
    -s_\delta & c_\delta
  \end{pmatrix}
  \begin{pmatrix}
    A'\\Z
  \end{pmatrix}_\text{mass}
  &
  \delta  &\approx 
  -\frac{m_Z^2}{m_{A'}^2-m_Z^2}\left(t_W \varepsilon - \kappa\right)
  \label{eq:diagonalization:2:mass:mixing}
\end{align}

\subsection{Interactions with Fermions}
\label{sec:interactions:with:matter}
The mass mixing causes the dark and Standard Model $Z$s to couple to linear combinations of the weak neutral and electromagnetic currents. We ignore dark\,$Z$ interactions to any low-mass dark sector states. If present, these simply reduce dark $Z$ branching ratios to visible states. Standard Model fermions are assumed to have zero charges under the U(1)$_\text{d}$ gauge group. The couplings of an \acro{SM} fermion $\psi$ to the \acro{SM} $Z$ and its dark counterpart have the form
\begin{align}
  \mathcal L &\supset 
  A'_\mu  \bar\psi \gamma^\mu \left(g^\text{V}+ g^\text{A}\gamma^5\right) \psi 
  + 
  \frac{g}{c_W} Z_\mu  \bar\psi \gamma^\mu \left(q_{Z}^\text{V}+ q_{Z}^\text{A}\gamma^5\right) \psi
  \ ,
  \label{eq:Z:Ap:couplings}
\end{align}
where $q_Z^\text{V,A}$ are the $\psi$ weak neutral charges.
In the absence of any mixing, the these charges take their Standard Model values,
\begin{align}
  q_{Z,\text{SM}}^\text{V} &= \frac{1}{2}T^3 - s_W^2 Q_{\text{EM}}
  &
  q_{Z,\text{SM}}^\text{A} &= -\frac{1}{2}T^3 
  \label{eq:Z:couplings:charges} \ ,
\end{align}
with $T^3=\pm 1/2$ according to the \acro{SU(2)}$_\text{L}$ weight of the left-chiral $\psi$ and $Q_\text{EM}$ is the $\psi$ electric charge. For example, an electron has $T^3 = -1/2$ and $Q_\text{EM} = -1$. 

\paragraph{Modified $Z$ couplings.}
The rotation to mass eigenstates \eqref{eq:diagonalization:2:mass:mixing} shifts the charges of fermions with respect to the Standard Model values,
\begin{align}
  q_{Z}^\text{V} &= 
  \left(c_\delta+ \varepsilon t_W s_\delta \right)
  q_{Z,\text{SM}}^\text{V} 
  + \frac{\varepsilon e}{g}c_W s_\delta\; Q_\text{EM}
  &
  q_{Z}^\text{A} &= 
  \left(c_\delta+ \varepsilon t_W s_\delta \right)q_{Z,\text{SM}}^\text{A} \ .
  \label{eq:qZnew:VA}
\end{align}
These corrections to the Standard Model charges are second order in $\varepsilon, \massmixing \ll 1$. 

\paragraph{Dark\,$Z$ couplings.} The $A'$ vector and axial couplings to $\psi$ are readily expressed in terms of the Standard Model $Z$ charges,
\begin{align}
  g^\text{V} &= \varepsilon e Q_\text{EM} - \frac{gr}{c_W} q_{Z,\text{SM}}^V
  &
  g^\text{A} &= \frac{-g r}{c_W} q_{Z,\text{SM}}^A 
  &
  r &\equiv
  \frac{m_Z^2\massmixing  + m_{A'}^2 \varepsilon t_W}{m_{A'}^2-m_Z^2} 
  \ .
  \label{eq:dark:Z:couplings}
\end{align}
These couplings arise at the linear order in $\varepsilon, \massmixing \ll 1$.
Unlike its coupling to the electric current, the dark $Z$ coupling to the weak neutral current depends on the relative masses of $A'$ and $Z$ in the factor $r$. 

\subsection{Interactions with Bosons}
\label{sec:interactions:with:bosons}

\begin{figure}[tb]
  % \centering
  \hspace{-2em}
  \includegraphics[width=1.06\textwidth]{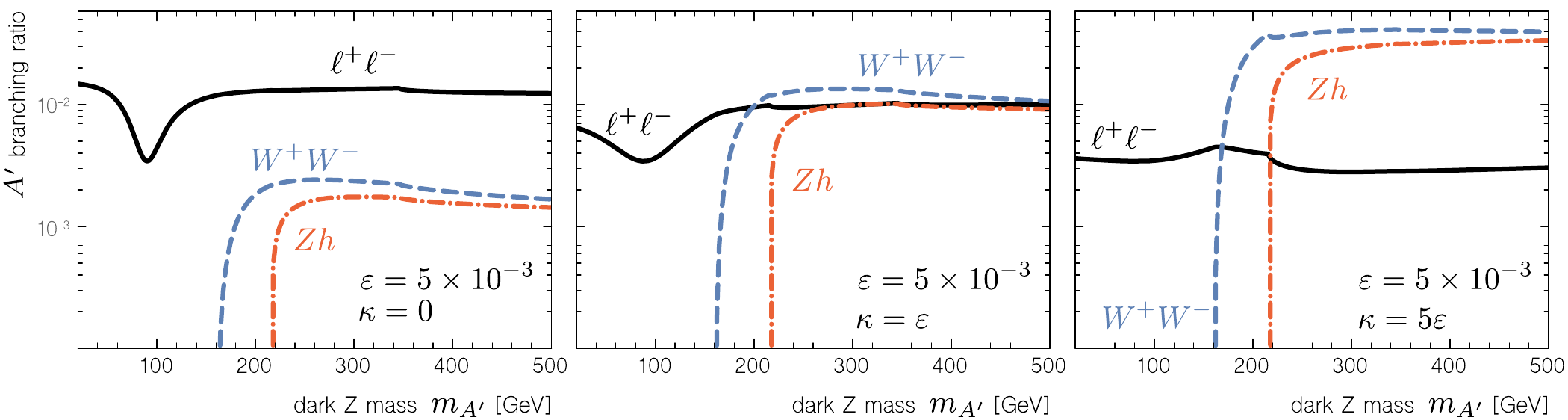}
  \caption{The branching ratios of $A'$ decays into dilepton pairs $\ell^+\ell^-$ and diboson pairs $W^+W^-$ and $Zh$ for different values of mass mixing $\kappa$ compared to kinetic mixing $\varepsilon$. For $\kappa \gtrsim \varepsilon$, the diboson branching ratios become large. For large $m_{A'}$, the diboson branching ratios converge as predicted by the Goldstone boson equivalence theorem.
  }
  \label{fig:branching:ratios}
\end{figure}

Dark\,$Z$ bosons that are at least twice as heavy as the $W$ boson may decay to $W^+W^-$. At higher masses, the $Zh$ channel is also accessible. 
In models with only kinetic mixing, the branching ratios for these channels are always at least an order of magnitude smaller than those for $\ell^+\ell^-$~ \cite{He:2017zzr}. However, the presence of mass mixing dramatically changes these results and the diboson final states can dominate, see Fig.~\ref{fig:branching:ratios}. To the best of our knowledge, the effect of these decay modes has not been addressed in past literature. 

The dark\,$Z$ inherits its coupling to $W^+W^-$ from its mixing with the $W^3$ component of the Standard Model $Z$, \eqref{eq:diagonalization:2:mass:mixing}. This coupling is thus $-s_\delta$ times the $ZW^+W^-$ coupling. The dramatic effect on the branching ratio for non-zero mass mixing, $\kappa \gtrsim \varepsilon$, can be understood as follows. Parametrically, $m^2_Z/m^2_{A^\prime}\ll 1$ in the region where the dark\,$Z$ decay to gauge bosons is kinematically allowed. In the unitary gauge, the couplings of the dark $Z$ to fermions and $W^+W^-$ are of the same order, $\mathcal O\left(\kappa\, m_Z^2/m_{A'}^2\right)$. However the spin sum for $W$ boson final state with momentum $k$ contributes an enhancement of $k_\mu k_\nu /m_W^2 \sim (m_{A'}/m_W)^2$ for each $W$ compared to the fermion final states. In the 't Hooft-Feynman gauge, the dark\,$Z$ decay to Goldstone modes is dominant in the limit $m_{A^\prime}\gg m_Z$. The dark\,$Z$ coupling to charged Goldstones is contained in the $H_\text{mix}$ kinetic term and is $\mathcal O\left(\kappa\right)$, without the additional suppression by $\left(m_Z^2/m_{A'}^2\right)$ compared to the fermion coupling.

The $A'Zh$ coupling depends on the mixing between the neutral \acro{CP}-even Higgs-like states in $H_\text{EW}$ and $H_\text{mix}$, an angle called $\alpha$ in the two-Higgs doublet literature. In our model, this mixing comes from integrating out the \acro{CP}-even state in $H_\text{d}$. In the large $m_{A'}/m_Z$ limit, the Goldstone boson equivalence theorem tells us that we may replace the $W^\pm$ and $Z$ final states by their eaten Goldstone bosons~\cite{Cornwall:1974km, Chanowitz:1985hj}. In the limit where the heavy states are decoupled, see Appendix~\ref{app:UV:model}, the eaten Goldstones and the low-energy \acro{CP}-even state are components of a \acro{SU(2)}$_L$ doublet and thus have the same couplings to the $A'$. Thus the branching ratios to $W^+W^-$ and $Zh$ approach each other at high dark\,~$Z$ mass, as shown in Fig.~\ref{fig:branching:ratios}.
There is no $A'ZZ$ coupling by \acro{SU(2)}$_L$ symmetry and no $Zhh$ coupling due to parity.

\section{Experimental Constraints and the ILC Reach}
\label{sec:exper}

\begin{figure}[tb]
  \centering
  % \hspace{-2em}
  \includegraphics[width=0.7\textwidth]{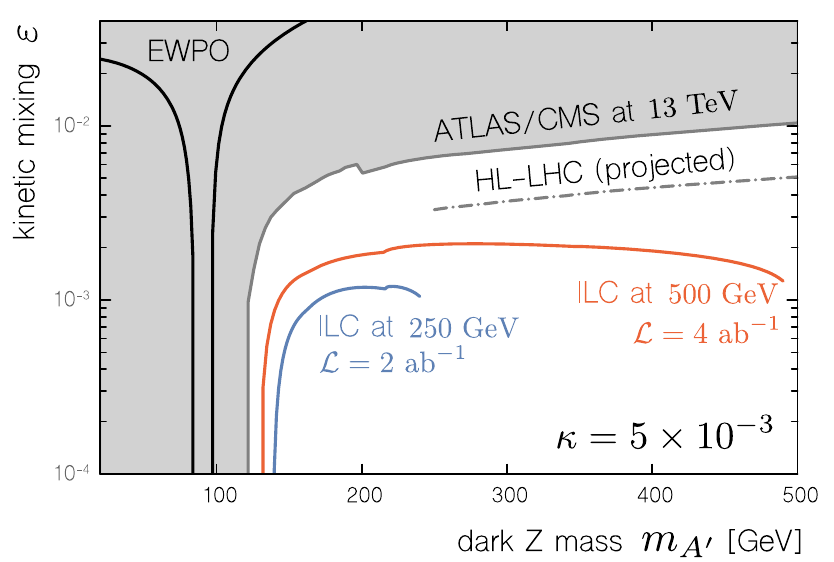}
  \caption{Discovery reach for a dimuon resonance search at the \acro{ILC} (colored) compared to bounds from Drell--Yan production at \acro{CMS}\cite{CMS:2019buh, CMS:2021ctt} and \acro{ATLAS}~\cite{ATLAS:2019erb}(gray shaded) for a benchmark value of the mass mixing, $\kappa$. The kink at $200~\text{GeV}$ is the transition between two separate recasted searches. We estimate the high-luminosity \acro{LHC} reach by rescaling the luminosity to 3000~fb$^{-1}$.  
  }
  \label{fig:reach:plot}
\end{figure}

This section explores the potential for a future $e^+e^-$ collider, such as the \acro{ILC}, to discover a dark\,$Z$. We begin by evaluating current constraints on the model parameter space from precision electroweak observables and the \acro{LHC},\footnote{Deep inelastic scattering observations constrain the presence of a dark $Z$ in a way that is agnostic to its decay modes. For visibly-decaying dark $Z$, the projected constraints from future electron--hadron colliders are subdominant to those from hadron colliders, both for unpolarized~\cite{Kribs:2020vyk} and polarized~\cite{Yan:2022npz} electron beams.} as well as estimating the likely reach of the \acro{HL-LHC}. We then study the process $e^+e^-\to \gamma A'$, followed by the decay $A^\prime\to \mu^+\mu^-$, at the \acro{ILC}, together with the relevant Standard Model backgrounds. The results for a benchmark value of the mass mixing parameter are summarized in Fig.~\ref{fig:reach:plot}. The main conclusion is that the \acro{ILC} can probe parameters well beyond the current bounds and the projected \acro{HL-LHC} sensitivity. Our study focuses on the mass range $m_{A'} > 20~\text{GeV}$.  Below this mass, the energy-frontier collider searches are less sensitive compared to fixed-target, beam-dump, and other low-energy experiments~\cite{Fabbrichesi:2020wbt,Battaglieri:2017aum,Alexander:2016aln}.\footnote{We note in passing that experiments using the beam dumps of a high-energy $e^+e^-$ collider such as the \acro{ILC} have impressive sensitivity to lighter dark photons~\cite{Kanemura:2015cxa,Asai:2021ehn}.}

\subsection{Electroweak Precision Observables}

The broad agreement of electroweak precision observables with their Standard Model values~\cite{Zyla:2020zbs} limits the amount of mixing between new gauge bosons and the Standard Model $Z$. 
Earlier work used this to constrain a dark photon model with pure kinetic mixing~\cite{Hook:2010tw, Curtin:2014cca}. We generalize those analyses to the case of both kinetic and mass mixing.
In our approximation, the tree-level shift in precision observables due to the dark\,$Z$ must be within the measurement errors. 
We assume that the observables are uncorrelated so that the covariance matrix is diagonal. These approximations are sufficient for our purpose of comparing the sensitivity of these measurements to direct searches at colliders.
We focus on the measurements made at the $Z$-pole and interpret the bounds as a constraint on the kinetic mixing $\varepsilon(m_{A'}, \kappa)$ as a function of the dark\,$Z$ mass and mass mixing parameter \eqref{eq:mass:mixing:definition}.

The leading-order predictions of the electroweak sector of the Standard Model contain three model parameters that need to be fixed by data. 
We establish a set of reference electroweak parameters: 
  the weak mixing angle $s_W^2 = \sin^2\theta_W$, 
  the electroweak vev $v^2$, 
  and the $Z$ coupling $g_Z^2 = g^2 + g'^2$. 
We fix the values of these parameters using the precise measurements of the fine structure constant, Fermi constant (from muon lifetime), and $Z$ mass:
\begin{align}
  \hat \alpha &= \frac{e^2}{4\pi} = \frac{g_Z^2 c_W^2 s_W^2}{4\pi}
  &
  \frac{\hat G_F}{\sqrt{2}} &= \frac{1}{2v^2}
  &
  \hat m_Z^2 &= \frac{g^2_Z v^2}{4} \ .
  \label{eq:EWPO:reference:expressions}
\end{align}
Quantities with a caret (e.g.~$\hat \alpha$) are measured quantities. One may invert \eqref{eq:EWPO:reference:expressions} to obtain expressions for the reference parameters in terms of the measured inputs. In the presence of new weakly-coupled physics, the reference parameters are shifted relative to their Standard Model values. For example, $g_Z^2 = (g_Z^2)^\text{SM}[1+\mathcal O(\xi)]$, where $\xi$ characterizes the new physics coupling. For our dark $Z$ model, a useful definition is
\begin{align}
  \xi \equiv \varepsilon t_W + \kappa \ll 1\ .
  \label{eq:xi}
\end{align}
The order parameter of electroweak symmetry breaking, $v^2$, is measured from the muon lifetime and is unchanged from the Standard Model:
\begin{align}
  v^2=\frac{1}{\sqrt{2}\hat G_F} \ .
  \label{eq:v2:EWPO}
\end{align}
In the dark\,$Z$ model, the Standard Model $Z$ boson mass is shifted, see \eqref{eq:mZ2:in:our:model}, which then shifts the reference value of the $Z$ coupling, 
\begin{align}
  g_Z^2(\xi, m_{A'}^2) = 
  4\sqrt{2}\, \hat G_F \hat m_Z^2
  \left[
  1 + 2\xi^2
  \frac{\hat m_{Z}^2}{m_{A'}^2 - \hat m_Z^2}
  + \mathcal O(\xi^4)
  \right] \ .
  \label{eq:gZ2:EWPO}
\end{align}
This shift in $g_Z^2$, in turn, shifts the reference expression for the weak mixing angle relative to the fine structure constant \eqref{eq:EWPO:reference:expressions},
\begin{align}
   s_W^2(\xi, m_{A'}^2) &= 
    \left(1-\hat D\right)
    \left[1-
      \frac{4\hat C \xi^2 }{\hat D(1-\hat D)} 
      \frac{\hat m_{Z}^2}{m_{A'}^2 - \hat m_Z^2}
      + \mathcal O(\xi^4)
    \right]
      &
    \hat C &\equiv \frac{\pi \hat\alpha}{\sqrt{2}\hat G_F \hat m_Z^2} 
    \ ,
    \label{eq:sW2:EWPO}
 \end{align}
 where $\hat D \equiv \sqrt{1-4\hat C}$. In our electroweak fit we work to leading non-trivial order in the dimensionless couplings, $\mathcal O(\xi^2)$.

\begin{table}
  \renewcommand{\arraystretch}{1} % spacing between rows
  \centering
  \begin{tabular}{ @{} r@{\hskip .75cm}r@{}l@{\hskip 1.75cm}l @{} } \toprule % @{} removes space
    Observable 
      & \multicolumn{3}{l}{
          Expression to second order in $\varepsilon$, $\kappa$
        }
    \\ 
    \hline
    \rule{0pt}{6ex} % manual spacer
    $Z$ width &
      $\displaystyle 
      \Gamma_Z = $
      &
      $\displaystyle 
      % \Gamma_Z = 
      \sum_f^{u,d,e,\nu} N_f \Gamma_f$
      &
      \textcolor{gray}{$
        \displaystyle \Gamma_f \equiv
        \frac{m_Z}{12\pi} \frac{g^2}{\cos^2\theta_W}
      \left[ 
        \left(q^V_{Z}\right)^2 + \left(q^A_{Z}\right)^2
      \right]_f
      $}
    \\
    \rule{0pt}{3ex} % manual spacer
    hadronic 
      &
      \multirow{2}{*}{
      $\displaystyle 
      \sigma_\text{had} =$}
      & \multirow{2}{*}{
      $\displaystyle 
      % \sigma_\text{had} = 
      \frac{12\pi}{m_Z^2}\frac{\Gamma_e\Gamma_\text{had}}{\Gamma_Z^2}$
      }
      &
      \textcolor{gray}{$
      \displaystyle \Gamma_\text{had} \equiv  3\left(2\Gamma_u+3\Gamma_\text{d}\right)
      $}
      \\
    cross section 
    \\
    \rule{0pt}{3ex} % manual spacer
    lepton ratio
      & $R_\ell =$
      & $%R_\ell = 
      \Gamma_\text{had}/\Gamma_e$
    \\
    \rule{0pt}{3ex} % manual spacer
    quark ratios
    & $R_q =$
    & $%R_q = 
    \Gamma_q/\Gamma_\text{had}$
    & for $q=u,d$
    \\
    \rule{0pt}{3ex} % manual spacer
    left--right & 
    \multirow{2}{*}{$
      \displaystyle A_{f} =$}
    &
    \multirow{2}{*}{$
      \displaystyle 
      % A_f = 
      \left.
      \frac{2q_{Z}^\text{V} q_{Z}^\text{A}}{\left(q_{Z}^\text{V}\right)^2+\left(q_{Z}^\text{A}\right)^2}
      \right|_{f}
      $}
      &
      for $f=e, u,d$
      \\
      asymmetry & & 
    \\
    \rule{0pt}{5ex} % manual spacer
    forward--backward & 
    \multirow{2}{*}{$
      \displaystyle A_{\text{FB},f} =$}
    &
    \multirow{2}{*}{$
      \displaystyle 
      % A_{\text{FB},f} = 
      \frac{3}{4}A_e A_f
      $}
      &
      for $f=e, u,d$
      \\
      asymmetry & & 
        % \\  
    \\
    \bottomrule
  \end{tabular}
  \caption{
    $Z$-pole precision observables. The effective $Z$ charges $q^\text{V,A}_Z$ are given in~\eqref{eq:qZnew:VA}. $N_f$ is the number of species for each fermion $f$ accessible in $Z$ decays; e.g.\ $N_u = 3\text{ colors }\times 2\text{ flavors}$, $N_{d} = 3\times 3$, $N_e = 3$, $N_\nu = 3$. $\Gamma_f$ is the associated partial width.
    \label{table:EWPO}
  }
\end{table}

We fix the mass-mixing parameter $\massmixing$ and find the maximum allowed kinetic mixing parameter $\varepsilon$ as a function of the dark $Z$ mass by performing a $\chi^2$ fit with respect to the remaining electroweak observables~\cite{Zyla:2020zbs}. For example, the $W$ mass is related to the $Z$ mass by $m_{W}^2 = c_W^2 m_Z^2$. This gives an expression for the $W$ mass in the dark $Z$ model,
\begin{align}
  m_W^2 &= \frac{1}{4}g_Z^2v^2 \left(1-s_W^2\right)
\end{align}
where the dependence on the new physics parameters comes from \eqref{eq:v2:EWPO}--\eqref{eq:sW2:EWPO}. The rest of the fit is set by the $Z$-pole observables in Table~\ref{table:EWPO}. Of these, only the $Z$ total width depends on $g_Z$. The remaining observables are ratios that depend on new physics parameters through the effective $Z$ charges, \eqref{eq:qZnew:VA}, as well as the new physics dependence of $s_W^2$ in \eqref{eq:sW2:EWPO} that appears in \eqref{eq:Z:couplings:charges}. 

\begin{figure}[tb]
  % \centering
  \hspace{-2.5em}
  \includegraphics[width=1.07\textwidth]{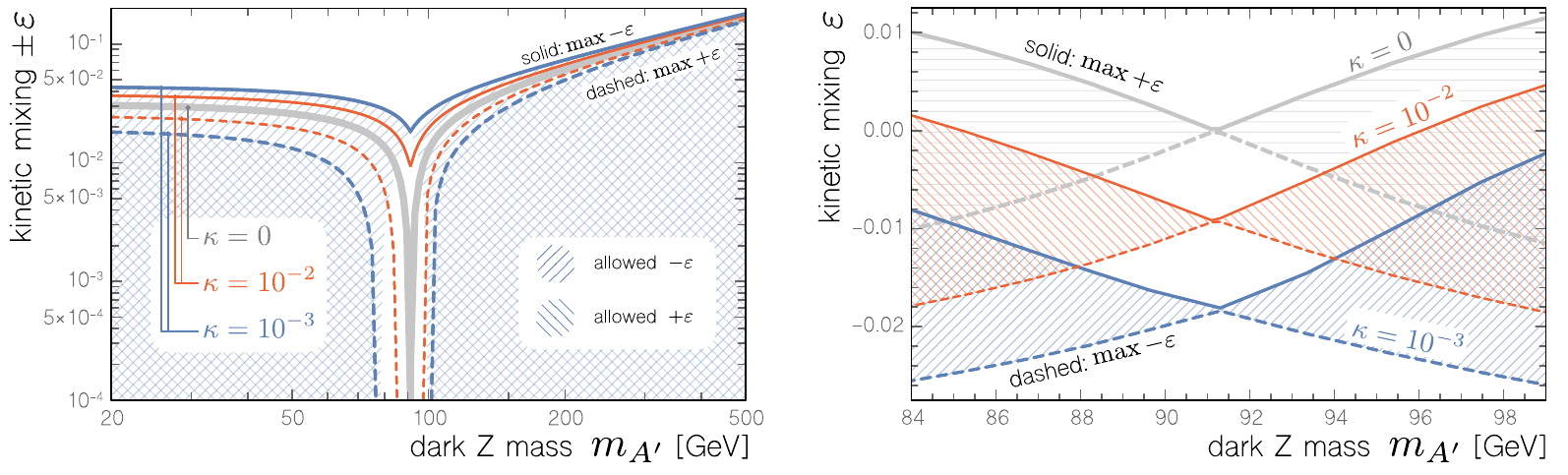}
  \caption{90\% confidence level constraints on the kinetic mixing parameter from electroweak precision observables as a function of dark\,$Z$ mass and for representative values of the mass mixing parameter, $\kappa$. 
  The relative sign between $\kappa$ and $\varepsilon$ matters so that bounds are asymmetric with respect to positive (solid line) and negative (dashed line) values of $\varepsilon$. \textsc{Left:} bounds on $|\varepsilon|$ for positive/negative values of $\varepsilon$ shown on a log scale. \textsc{Right:} bounds on a linear scale, showing the allowed region (hashed) near the $Z$-mass where mixing is large, \eqref{eq:diagonalization:2:mass:mixing}.
  }
  \label{fig:reach:EWPO}
\end{figure}

We present the 90\% confidence precision electroweak constraints on the kinetic mixing in Fig.~\ref{fig:reach:EWPO} for fixed values of the mass mixing, $\kappa$. For the case of dark photon, $\kappa=0$, our results are consistent with Refs.~\cite{Hook:2010tw, Curtin:2014cca}. In the presence of mass mixing, the constraints depend on relative sign of $\varepsilon$ and $\kappa$; see \leqn{eq:qZnew:VA} and~\leqn{eq:xi}. The constraint becomes weaker when $\varepsilon$ and $\kappa$ have the same sign, and stronger otherwise. The dependence on the dark\,$Z$ mass is the same as in the case of dark photon: the constraint is strongest for $m_{A^\prime}\approx m_Z$, and weakens rapidly outside this region.   

\subsection{Drell--Yan Production at Hadron Colliders}

The Large Hadron Collider produces on-shell dark\,$Z$ bosons through the Drell--Yan process, $q\bar q \to A'$. To evaluate the \acro{LHC} reach, we focus on the dimuon resonance search, looking for decays $A^\prime \to \mu\bar \mu$. A similar reach is expected in the di-electron channel, while the quark and tau channels have much lower sensitivity due to large \acro{QCD} backgrounds. In the region of parameter space where mass mixing dominates, the branching ratio to muons is suppressed and decays to $W^+W^-$ and $Zh$ become dominant; see Sec.~\ref{sec:interactions:with:bosons}. The \acro{LHC} reach in that region can be improved by searching for a resonance in boson-pair final states. We defer this analysis to future work.  

The \acro{LHC} collaborations report limits on the production cross section on resonances decaying into a dilepton pair\cite{CMS:2019buh, CMS:2021ctt, ATLAS:2019erb}. We recast these limits into bounds on our model parameters by computing the Drell--Yan cross section using the differential parton luminosity\footnote{We use the parton luminosities ${d\mathcal L_{a\bar a}}/{d\hat s}$ provided by the \emph{Mathematica} program \texttt{ManeParse}~\cite{Clark:2016jgm} and based on the \acro{nCTEQ15} parton distribution functions~\cite{Kovarik:2015cma}.} for partons $a$ and $b$,
\begin{align}
  \frac{d\mathcal L_{ab}}{d\hat s}
  &=
  \frac{1}{s}
  \frac{1}{1+\delta_{ab}}
  \int_\tau^1 \frac{dx}{x} 
    f_a\left(x,\sqrt{\hat s}\right)
    f_b\left(\frac{\tau}{x},\sqrt{\hat s}\right)
    + 
    (a\leftrightarrow b) \ ,
\end{align}
where $\tau = \hat s/s$ is the ratio of the partonic to the proton center-of-mass collision energies-squared. The $(1+\delta_{ab})^{-1}$ factor accounts for double counting identical initial partons, though this is not relevant for Drell--Yan production of the dark\,$Z$ since the quark and antiquark are distinguishable. The proton--proton cross section $\sigma_\text{pp}$ is related to the partonic cross section $\hat \sigma_{ab}$ as
\begin{align}
  \sigma_\text{pp} = \sum_{ab} \int 
  \frac{d\hat s}{\hat s}
  \frac{d\mathcal L_{ab}}{d\hat s}
  \left(\hat s \hat \sigma_{ab}\right) 
  = 
  \sum_f^\text{quarks} 
  \frac{d\mathcal L_{f\bar f}}{d\hat s} 
  \hat\sigma_{f\bar f}(m_{A'}^2) 
  \ ,
  \label{eq:parton:luminosity}
\end{align}
where the right-hand side sums over quark flavors $f$ with associated antiquark of flavor $\bar f$.
\begin{figure}[tb]
  \centering
%   \hspace{-2em}
  \includegraphics[width=0.7\textwidth]{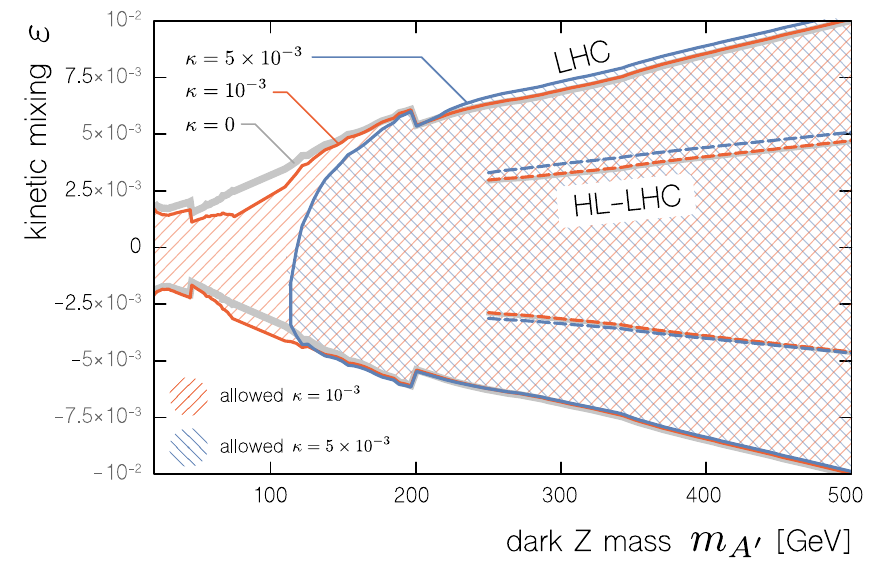}
  \caption{Effect of different values of $\kappa$ on the \acro{LHC} constraints. The vertical axis is on a linear scale to show how the relative sign of $\kappa$ and $\varepsilon$ affects the size of the coupling; see  \eqref{eq:dark:Z:couplings}. 
  }
  \label{fig:reach:plot:right}
\end{figure}

We ignore kinematic thresholds and use the narrow-width approximation in the partonic cross section $\hat \sigma$. The flavor dependence shows up in the production cross section, which is proportional to the square of the effective couplings,
\begin{align}
  \hat\sigma_{f\bar f}(m_{A'}^2) 
  &= 
  \left[ (g^\text{V}_f)^2 + (g^\text{A}_f)^2\right] 
  \tilde\sigma(m_{A'}^2)
  \,.
  \label{eq:sigma:hat:qV:qA}
\end{align}
The couplings $g^\text{V}$ and $g^\text{A}$ are given in~\leqn{eq:dark:Z:couplings}, and the rescaled cross section $\tilde{\sigma}$ is flavor-universal. 

We recast the \acro{LHC} dilepton results by setting $\sigma_\text{pp}\qty(m_{A'}, \varepsilon, \massmixing) < \sigma_\text{LHC}$, where $\sigma_\text{LHC}$ is the upper limit set by \acro{LHC} analyses. This straightforwardly produces the bounds on the effective theory parameters $\varepsilon$ and $\kappa$ shown in Fig.~\ref{fig:reach:plot:right}. In the case of pure kinetic mixing, our results are in agreement with the bounds reported by the \acro{CMS} collaboration~\cite{CMS:2019buh}. We also extrapolate the current \acro{LHC} bounds to estimate the reach of the full \acro{HL-LHC} data set with 3000~fb$^{-1}$, assuming that statistical errors dominate so that the sensitivity to new physics cross section scales as the square root of the integrated luminosity.

\subsection{Resonance Search at the ILC}

\begin{figure}[tb]
  % \centering
  \hspace{-3em}
  \includegraphics[width=1.07\textwidth]{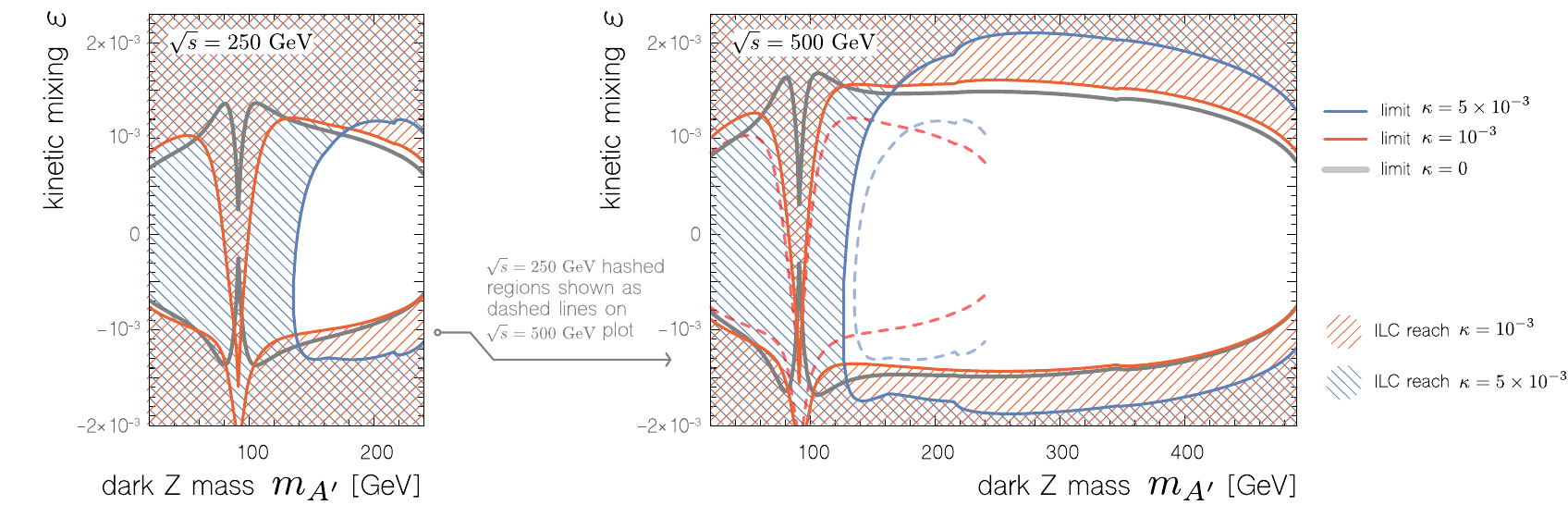}
  \caption{Dimuon resonance discovery reach at the \acro{ILC} (${\mathcal L}_{\rm int}=2$~ab$^{-1}$ at 250 GeV and 4 ab$^{-1}$ at 500 GeV). Stronger reach corresponds to shaded regions closer to $\varepsilon=0$. We plot both positive and negative values of the kinetic mixing, $\varepsilon$, for different values of the mass mixing, $\kappa$. The relative sign of $\kappa$ and $\varepsilon$ is significant, as seen in the couplings \eqref{eq:dark:Z:couplings} and in the asymmetry of these plots for $\varepsilon\to-\varepsilon$. 
For $\kappa = 10^{-3}$ (\textcolor{red}{red lines}) at masses near 120 GeV, the $\varepsilon<0$ branch has a stronger reach than $\kappa=0$, whereas for $\varepsilon>0$ the reach is weaker than $\kappa=0$. For larger $\kappa$ (e.g.~\textcolor{blue}{blue lines}), the mass mixing is so large that it dominates over the kinetic mixing. }
  \label{fig:ILC:reach:varyingkappa}
\end{figure} 

We examine the reach for the \acro{ILC} to discover the dark\,$Z$ in the radiative return channel $e^+ e^- \to \gamma A' \to \gamma \mu^+ \mu^-$. 
The signal rate is strongly enhanced when the photon is collinear with the beam and hence unobservable.\footnote{He~et al.~estimate the discovery reach for a model with pure kinetic mixing at a future lepton collider with cuts that require the photon to be observable~\cite{He:2017zzr}. This is overly conservative because a high-mass dimuon resonance is a clean signal whether or not the photon is tagged. } 
We thus calculate the signal as on-shell $e^+e^-\to A'$ production at center-of-mass squared energy $x s$ in the {equivalent-particle approximation} where the beam of energy $E = \sqrt{s}/2$ contains a distribution of lower-energy $e^\pm$ particles with energy $xE<E$ that account for the unobserved collinear photon of energy $E_\gamma = (1-x)E$~\cite{vonWeizsacker:1934nji,Williams:1935dka, Chen:1975sh}. (See Ref.~\cite{Perelstein:2010hh} for a pedagogical introduction.) At the leading order in  \acro{QED} perturbation theory, the probability of finding an $e^\pm$ with energy $xE$ in the beam is
\begin{align}
  f_e(x)\,dx&= \frac{\alpha}{2\pi} \frac{1+x^2}{1-x} \ln\frac{\sqrt{s}}{m_e}\,dx \ .
\end{align}
The total signal cross section for $e^+e^-\to \gamma(A'\to\mu^+\mu^-)$ is readily expressed in terms of the cross section for two-to-one production of the dark $Z$:
\begin{align}
  \sigma &= \frac{\alpha}{\pi}\ln \frac{\sqrt{s}}{m_e}
            \int_0^1 dx\,\frac{1+x^2}{1-x} \hat\sigma(e^+e^-\to A')
            \, \text{Br}(A'\to \mu^+\mu^-) \ .
\end{align}
In the narrow-width approximation, $\hat\sigma \propto \delta(s'-m_{A'}^2)$, where $s^\prime = x s$ is the center-of-mass energy-squared of the ``partonic" collision. The cross section for background events mediated by photons or \acro{SM} $Z$ bosons follows a similar formula, except that narrow-width approximation does not hold since we will be searching in regions where the photon or $Z$ are off-shell (unless $m_{A'} \sim m_Z$). The dark\,$Z$ appears as a narrow peak in the dimuon invariant mass distribution. We estimate the signal significance by computing $S=N_\text{sig}/\sqrt{N_\text{bg}}$, where $N_\text{bg}$ is the number of background events in the mass window centered at $m_{A^\prime}$. The intrinsic width of the dark\,$Z$ is very small in the region of interest due to its small couplings. Thus the size of the mass window is controlled by muon momentum resolution. We use $\Delta\qty(1/p_T) \sim 10^{-5} \text{ GeV}^{-1}$~\cite{Behnke:2013lya} in our estimates.

The \acro{ILC} reach at $2\sigma$ confidence for a benchmark value of $\kappa = 5\times 10^{-3}$ is shown as the colored lines in Fig.~\ref{fig:reach:plot}. This analysis demonstrates that the \acro{ILC} can serve as the discovery machine for this type of new physics. The \acro{ILC} reach significantly exceeds both the current limits and the extrapolated sensitivity of the high-luminosity \acro{LHC} in the kinematically accessible regions. Fig.~\ref{fig:ILC:reach:varyingkappa} presents the \acro{ILC} reach for different values of the mass mixing $\kappa$. We note that for $\kappa=0$, our results are in agreement with the previous estimates of the dark photon reach in Refs.~\cite{Karliner:2015tga,Berggren:2021lan}.

Our analysis assumes unpolarized \acro{ILC} beams. Depending on the model parameters, the couplings of the dark\,$Z$ can depend strongly on the lepton chirality, while the background does not have strong polarization dependence. In such models, the appropriate beam polarization may further enhance the \acro{ILC} reach. However, since the chiral structure of the couplings is not known {\it a priori}, running with unpolarized beams is overall the preferred search strategy. If a dark\,$Z$ is discovered, polarized beams can be used to measure its chiral couplings; we explore this in the next section.     

We also note that, unlike the \acro{LHC}, hadronic decays of the dark\,$Z$ do not suffer from large backgrounds at a lepton collider, and can be included to further improve the \acro{ILC} reach. We leave this analysis for future work. 

\section{\texorpdfstring{Precision Measurements at the Dark\,$Z$ Pole}{Precision Measurements at the Dark Z Pole}}
\label{sec:Zpole:Obs}

Electron--positron colliders such as the \acro{ILC} are an ideal laboratory to determine the chiral couplings of a dark\,$Z$ resonance. This program is analogous to the precision measurements of the electroweak gauge bosons at \acro{LEP} and \acro{SLC}.  
To illustrate this capability, we consider the benchmark point %(\acro{BP})
\begin{align}
  m_{A'} &= 400~\text{GeV}
  &
  \varepsilon &= 5 \times 10^{-3}
  &
  % s_\beta^2 
  \kappa
  &= 5 \times 10^{-3} \ .
  \label{eq:benchmark:point}
\end{align}
A dark\,$Z$ at this benchmark would be already discovered at the \acro{HL-LHC}, see Fig.~\ref{fig:reach:plot}. Models with slightly smaller couplings may evade the \acro{HL-LHC} and be first discovered at a 500~GeV \acro{ILC}. In either case, an electron--positron collider would provide the first opportunity to measure the dark $Z$ couplings to fermions with high precision, and to determine their chiral structure which is crucial for model discrimination. 

\begin{figure}[tb]
  % \centering
  \hspace{-2em}
  \includegraphics[width=1.05\textwidth]{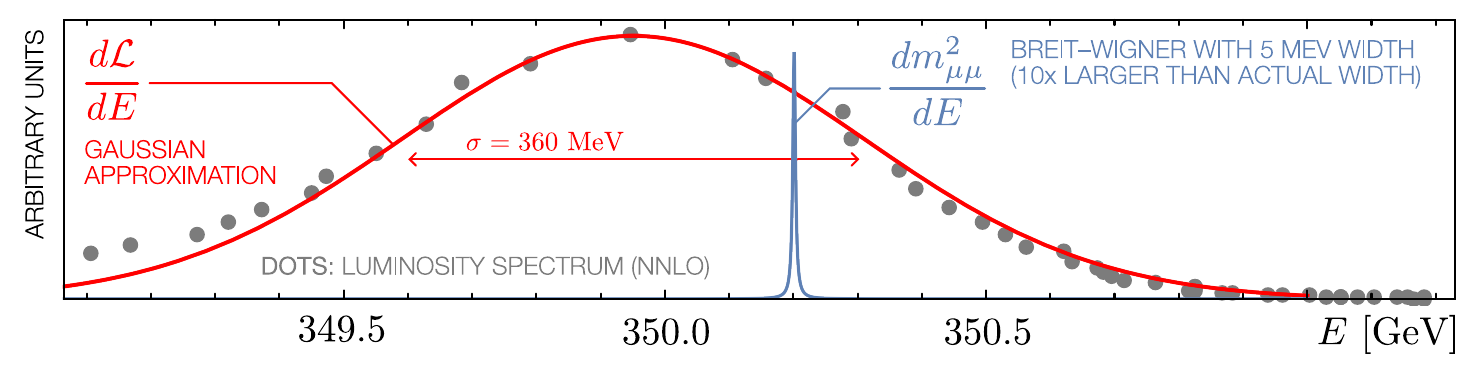}
  \caption{\acro{ILC} beam luminosity spectrum, for nominal center-of-mass energy $\sqrt{s}=350.5$~GeV. Dots are the simulation results from Fig.~2 of Ref.~\cite{Simon:2014hna}, while the red line is a Gaussian fit. The Gaussian width is $\sigma = 360~\text{MeV}$.
  We compare to a hypothetical Breit--Wigner distribution for a dark~$Z$ with mass $m_{A'}=350.2~\text{GeV}$. We choose a dark~$Z$ width $\Gamma_{A'} = 5~\text{MeV}$; this is $10\times$ larger than the expectation for the benchmark theory so that the width is visible in the plot.
  Unlike typical models of $\mathcal O(\text{TeV})$ spin-1 gauge bosons, the luminosity spectrum spread is much broader than the dark\,$Z$ width.
  Vertical axis normalization is arbitrary.
  }
  \label{fig:lumi:width}
\end{figure}

The mass of the dark\,$Z$ will be measured experimentally at both the \acro{HL-LHC} and through the radiative return production in the 500~GeV \acro{ILC} run discussed above. The couplings of the dark\,$Z$ may be measured efficiently in a relatively short period of time with a dedicated \acro{ILC} run at a center-of-mass energy $\sqrt{s}=m_{A^\prime}$. In this section, we estimate the potential of this run to measure the dark\,$Z$ couplings to charged leptons. We study the channel $e^+e^-\to A^\prime\to \mu^+\mu^-$ assuming flavor-universal couplings; this assumption is automatically satisfied in our underlying model. The left- and right-handed couplings are defined by 
\begin{align}
g_\text{R,L} &= \frac{1}{2}\left(g^\text{V}\pm g^\text{A}\right)
&
g_\text{R} &= -2.1 \times 10^{-3}
&
g_\text{L} &= -9.3 \times 10^{-4}
\,,   
\label{eq:gLR}
\end{align}
and are related to the underlying theory parameters in~\eqref{eq:dark:Z:couplings}; we give numerical values for the benchmark point \eqref{eq:benchmark:point}. To measure these couplings, we consider three observables:  
\begin{itemize}
  \item The number of events, $N_\text{prod}$  
  \item The forward--backward asymmetry, $A_\text{FB}$ 
  \item The left--right asymmetry for polarized beams, $A_\text{LR}$.
\end{itemize}
The expressions for the two asymmetries are well-known and are summarized in Appendices~\ref{app:unpolarized:observables} and~\ref{app:polarized:beams}. The number of events is given by
\begin{align}
N_\text{prod}&=
\int d{\sqrt{s}} \; \frac{d\mathcal{L}}{d\sqrt{s}} \sigma_{\rm prod}, 
\label{eq:Nprod}
\\
\sigma_\text{prod} &=
\frac{\pi}{2} \qty[(1 - P_{e^-})\qty(1+P_{e^+}) g_L^2
+ (1 + P_{e^-})\qty(1 - P_{e^+}) g_R^2] \delta\qty(s - m_{A'}^2).
\label{eq:sigma_prod}
\end{align} 
Here $P_{e^-}$ and $P_{e^+}$ denote polarizations of the electron and positron beams, respectively and $d{\mathcal{L}}/d{\sqrt{s}}$ is the differential luminosity of the colliding beams at the \acro{ILC}.
The somewhat unusual formula for $N_\text{prod}$ is due to the fact that intrinsic width of the $A^\prime$ is extremely small, $\Gamma_{A^\prime}\sim 10^{-4}$~GeV for our benchmark parameters \eqref{eq:benchmark:point}. The beam luminosity spectrum of the \acro{ILC} in the vicinity of the nominal center-of-mass energy can be modeled as a Gaussian with a width of order $\mathcal O\left(1~\text{GeV}\right)$~\cite{Simon:2014hna}, much larger than $\Gamma_{A^\prime}$, as illustrated in~Fig.~\ref{fig:lumi:width}. Treating the Breit--Wigner resonance as a $\delta$-function yields \leqn{eq:sigma_prod}. 

\begin{figure}[tb]
  % \centering
  \hspace{-2em}
  \includegraphics[width=1.06\textwidth]{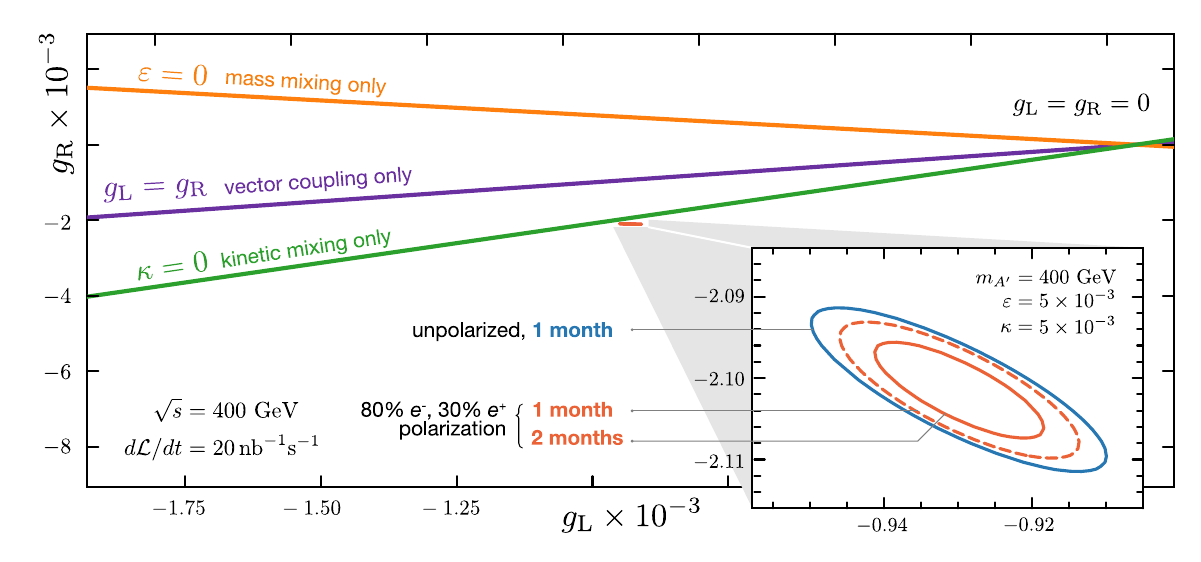}
  \caption{\acro{ILC} measurement of the dark\,$Z$ chiral couplings to leptons. 
  The ellipses correspond to 95\% confidence level, assuming that the best-fit values of the couplings match the underlying model predictions and that statistical errors dominate. The lines represent alternative models of a narrow spin-1 resonance that can be tested by this measurement. 
  }
  \label{fig:ellipse}
\end{figure}

The prediction for $N_\text{prod}$ depends sensitively on the relative location of the Breit--Wigner resonance with respect to the peak of the luminosity spectrum. To obtain a precise constraint on the couplings from a measurement of $N_\text{prod}$, this relative location must be known to a precision $\ll 1$~GeV. This is a non-trivial requirement. The position of the peak of the luminosity spectrum will be known at the level of $10^{-4}E_\text{beam} \sim 40$~MeV~\cite{Muchnoi:2008bx}, which is sufficient. However, prior to the dedicated on-resonance run of the \acro{ILC}, the mass of the $A^\prime$ will only be measured as the location of the peak in the dimuon invariant mass distribution, observed either at the \acro{HL-LHC} or in the radiative-return sample collected at the 500~GeV \acro{ILC}. In either case, the precision of the mass measurement is limited by the muon momentum resolution, which is of order GeV. A much more precise knowledge of the location of the Breit--Wigner peak is required.  

To address this problem, we propose that the on-resonance run of the \acro{ILC} should begin with an {\it inverse lineshape scan} in which the ``peak'' center-of-mass energy---the location of the peak of luminosity spectrum---is varied in small steps within the $A^\prime$ mass window inferred from the \acro{HL-LHC}/\acro{ILC}-500 data. For example, our benchmark point \eqref{eq:benchmark:point} motivates $\sqrt{s}=399$--$401$ GeV.  At each beam energy step, measure the $A^\prime$ production cross section. Since the luminosity spectrum is independently known, the $m_A^{\prime}$ can be inferred from the measured cross sections. We estimate that with sufficient statistics at each energy step, this method can provide an error on $m_A^{\prime}$ comparable to the peak beam energy uncertainty, $\sim 40$~MeV. After performing the inverse lineshape scan, the peak center-of-mass energy can be chosen to match the measured $m_A^{\prime}$ and maximize the dark-$Z$ production rate.

Fig.~\ref{fig:ellipse} shows the expected statistical errors of the measurement of $g_\text{L}$ and $g_\text{R}$ with one-month and two-month runs on the dark $Z$ resonance. The ellipses correspond to 95\% confidence level measurement, assuming that the best-fit values of the couplings match the underlying model predictions. Also shown are the predictions of three alternative theoretical models of a vector resonance: pure kinetic mixing\footnote{The benchmark point is close to the $\kappa = 0$ line even though $\kappa \sim \varepsilon$ because the contribution of $\kappa$ to the chiral couplings is suppressed by $ m_Z^2/m_{A'}^2\sim 0.05$.} ($\kappa=0$), mass mixing only ($\epsilon=0$), a parity-conserving ($g_\text{L}=g_\text{R}$). It is clear from the figure that the measurement of the chiral couplings at the \acro{ILC} can rule out all three possibilities, establishing parity violation and proving that the underlying dark\,$Z$ model has both kinetic and mass mixing. In contrast, the \acro{LHC} can only constrain the combination $\sqrt{g_\text{L}^2+g_\text{R}^2}$, which is not sufficient to discriminate between alternative models.

\begin{table}[t]
  \renewcommand{\arraystretch}{1.3} % spacing between rows
  \centering
  % \begin{tabular}{ @{} llll @{} } 
  \begin{tabular}{ @{} r@{\hskip 1cm}l@{\hskip 1cm}l @{} } \toprule % @{} removes space
    & $\phantom{+}g_\text{L} \times 10^{5}$ 
    & $\phantom{+}g_\text{R} \times 10^{5}$ 
    \\ \hline
    Benchmark values & $-93.0$ & $-210$ \\ \hline
    1 month (unpolarized) & $\pm 4.0 \pm 0.3$ & $\pm2.0 \pm 0.8$ \\ \hline
    1 month (polarized) & $\pm 3.3 \pm 0.3$ & $\pm1.7 \pm 0.7$ \\ %\hline
    2 months (polarized) & $\pm 2.3 \pm 0.3$ & $\pm0.85 \pm 0.7$ %\\ \hline
    \\ \bottomrule
  \end{tabular}
  \caption{
  Chiral couplings $g_\text{L,R}$ for the benchmark values in \eqref{eq:gLR} and the statistical (first number) and systematic (second number) uncertainties for different \acro{ILC} run times and beam polarizations. Polarized beams assume 80\% and $-30$\% polarizations in $e^-$ and $e^+$ beams, respectively.
  }
  \label{tab:precisionILC}
\end{table}

The constraint on the couplings inferred from the total number of events $N_\text{prod}$ is subject to systematic errors from the uncertainties in luminosity and peak beam energy and from the uncertainty in the measurement of the mass $m_{A^\prime}$. The mass is measured through the inverse lineshape scan. The expected error is of the same order as that in the peak beam energy, $\delta m_{A^\prime} \approx 40$~MeV. The expected overall luminosity uncertainty is $(\Delta \mathcal L/\mathcal L)\sim 4\times 10^{-3}$~\cite{BozovicJelisavcic:2013lni}. We estimate that these uncertainties affect the coupling determination at a level that is, at most, comparable to the statistical errors shown in Fig.~\ref{fig:ellipse}. We do not include uncertainties in the shape of the beam luminosity spectrum. The measurement of the left-right asymmetry $A_\text{LR}$ is subject to beam polarization uncertainties. \acro{ILC} polarimeters can achieve a precision of $\sim0.25$\%~\cite{Boogert:2009ir}. The effect of this uncertainty on the coupling measurement is negligible. We collect our estimates of the statistical and combined systematic errors of the chiral coupling measurements in Table~\ref{tab:precisionILC}. We conclude that a dedicated run of the \acro{ILC} at the dark\,$Z$ pole will yield percent-level measurement of $g_\text{L}$ and $g_\text{R}$, providing crucial information for discrimination among alternative models of the resonance.

\section{Conclusions}
\label{sec:conc}

A next-generation electron--positron collider, such as the \acro{ILC}, offers unique opportunities to search for and characterize new physics beyond the Standard Model. In this paper, we study a model of a new Abelian gauge boson, the dark\,$Z$. While \acro{SM} fermions are not directly charged under the new gauge group, the dark\,$Z$ can interact with the \acro{SM} through both kinetic and mass mixing. If the dark\,$Z$ mass is above 10~GeV or so, energy-frontier colliders offer the best way to search for this particle. We studied the current constraints on this particle from precision electroweak measurements and direct searches at the \acro{LHC}. This extends the previous work on the dark photon model that couples to the \acro{SM} only via kinetic mixing to the more general case when mass mixing is also present. We evaluate the reach of the \acro{ILC}, running at 250 and 500~GeV, in the dark\,$Z$ parameter space. The \acro{ILC} reach covers a large part of the parameter space unconstrained by current experiments and not accessible at the \acro{HL-LHC}; see Fig.~\ref{fig:reach:plot}. Thus, the \acro{ILC} can serve a discovery machine in this new physics scenario. If a dark\,$Z$ is discovered, the \acro{ILC} offers a unique opportunity to precisely measure its couplings to the \acro{SM} fermions, including their chiral structure. We demonstrated that a short dedicated run at the center-of-mass energy matching the dark\,$Z$ mass will achieve percent-level measurement of the chiral couplings of the dark\,$Z$ to leptons. Such a measurement can play a crucial role in model discrimination; this is illustrated in Fig.~\ref{fig:ellipse}.

The idea of using on-resonance electron-positron collisions to measure the properties of a massive vector boson is, of course, not new: \acro{LEP-1} and \acro{SLC/SLD} used it to understand the Standard Model $Z$ with unprecedented precision. A new twist in the case of dark\,$Z$ is its extremely small intrinsic width, which is much smaller than the width of the beam luminosity spectrum as well as the energy/momentum resolution of the detector for all visible decay products. In this situation, precise measurement of the mass and couplings of the resonance hinges on understanding the beam luminosity spectrum. Note that the small intrinsic width of the dark\,$Z$ is directly related to the smallness of its couplings to \acro{SM} fermions, which is in turn necessary for a sub-TeV resonance to avoid precision electroweak and \acro{LHC} constraints. Independently of the underlying model, any resonance compatible with current data and kinematically accessible at the \acro{ILC} will necessarily be very narrow. This motivates further studies of on-resonance production of such narrow states at the \acro{ILC} and other proposed lepton colliders.

\section*{Acknowledgments}

We thank
Mikael Berggren,
Philip Ilten,
Gopolang Mohlabeng,
Michael Peskin, 
and
Yotam Soreq
for insightful discussions. 
\acro{MP} and \acro{YCS} are supported by the National Science
Foundation (\acro{NSF}) grants \acro{PHY-1719877} and \acro{PHY-2014071}. \acro{PT} is supported by an \acro{NSF CAREER} award. 
\acro{PT} thanks the Aspen Center for Physics (\acro{NSF} grant \acro{\#1066293}) and Kavli Institute for Theoretical Physics (\acro{NSF} grant \acro{PHY-1748958}) for hospitality during a period when part of this work was completed. 

%% Appendices
\appendix

\section{Simplified Model and Feynman Rules}
\label{app:model:summaru}

Our effective dark\,$Z$ model is parameterized by the dark\,$Z$ mass, $m_{A'}$, its kinetic mixing to hypercharge, $\varepsilon$, and a dimensionless $Z$--dark\,$Z$ mass mixing parameter, $\massmixing$. These are related to a benchmark ultraviolet model in Section~\ref{sec:simplified:model}. We use the shorthand $s_W=\sin\theta_W$, and similarly for cosine, tangent, and the massive gauge boson mixing angle $\delta$, \eqref{eq:diagonalization:2:mass:mixing}.

\paragraph{Mass Spectrum.}
The mass spectrum comes from diagonalizing the mass terms \eqref{eq:mass:terms:from:vevs}. Assuming $m_{A'} > m_Z$, the $Z$ boson mass to second order in the new physics parameters is 
\begin{align}
  m_Z^2 
  &=
  M_Z(v)^2
      \left[1 -
        \frac{2M_Z(v)^2}{m_{A'}^2 -M_Z(v)^2}
        (\varepsilon t_W + \massmixing)^2
      \right]
 &
 M_Z(v)^2
 &\equiv
 \frac{g^2v^2}{4c^2_W}
 \ ,
 \label{eq:mZ2:in:our:model}
\end{align}
where $v^2 = (246~\text{GeV})^2$ is the order parameter for electroweak symmetry breaking. For completeness, the dark\,$Z$ mass is related to the benchmark ultraviolet model parameters to leading order in the new physics parameters by
\begin{align}
  m_{A'}^2
  &=
  M_{A'}^2
  + 
  \frac{2M_Z(v)^4}{M_{A'}^2 -M_Z(v)^2}
        (\varepsilon t_W + \massmixing)^2
  &
  M_{A'}^2 &= 
  g_\text{d}^2 v_\text{d}^2
  + \frac{\massmixing}{2 q_\text{d}} \frac{gg_\text{d}}{c_W} v^2
  \ .
\end{align}

\paragraph{Feynman Rules.}
The interactions with Standard Model fermions $\psi$ are described by% the Feynman rules
\begin{align}
  \vcenter{
    \hbox{\includegraphics[width=.15\textwidth]{{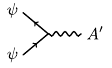}}}
    }
  &=
  i \gamma^\mu\left(g^\text{V}+g^\text{A}\gamma^5\right) 
  &
  \vcenter{
    \hbox{\includegraphics[width=.15\textwidth]{{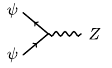}}}
    }
  &=
  i \frac{g}{c_W}\gamma^\mu\left(q_Z^\text{V}+q_Z^\text{A}\gamma^5\right) 
  \ .
\end{align}
The effective couplings and charges to leading non-trivial order in $\varepsilon$ and $\massmixing$ are
\begin{align}
  g^\text{V} &= \varepsilon e Q_\text{EM} 
  - \frac{g}{c_W} \frac{m_Z^2\massmixing + m_{A'}^2 \varepsilon t_W}{m_{A'}^2-m_Z^2} \left(\frac{1}{2}T^3 - s^2_W Q_\text{EM}\right)
  \\
  g^\text{A}&= 
  \frac{g}{c_W} \frac{m_Z^2\massmixing + m_{A'}^2 \varepsilon t_W}{m_{A'}^2-m_Z^2} 
  \frac{1}{2}T^3
  \\
  q_{Z}^\text{V} &= \phantom{+}
  \left(c_\delta+ \varepsilon t_W s_\delta \right)
  \left(\frac{1}{2}T^3 - s^2_W Q_\text{EM}\right)
  + \frac{\varepsilon e}{g}c_W s_\delta\; Q_\text{EM}
  \\
  q_{Z}^\text{A} &= -
  \left(c_\delta+ \varepsilon t_W s_\delta \right)
  \frac{1}{2}T^3 \ ,
\end{align}
where
$T^3=\pm 1/2$ according to the \acro{SU(2)}$_\text{L}$ weight of the left-chiral $\psi$ and $Q_\text{EM}$ is the $\psi$ electric charge. 
Additional interactions between the $A'$, Higgs, and electroweak gauge bosons are not used in this study. The effects are well approximated using the Goldstone boson equivalence theorem, as discussed in Section~\ref{sec:interactions:with:bosons}. See Ref.~\cite{He:2017zzr} for explicit expressions in the pure kinetic mixing case. 

\section{UV Model and Decoupling Limit}
\label{app:UV:model} 

In Section~\ref{sec:symmetry:breaking:extended:higgs} we derive our three-parameter model for mass and kinetic mixing as the low-energy theory of an extended Higgs sector. This appendix addresses some model-building aspects of the \acro{UV} model and connects to known results in the two-Higgs doublet literature.

\paragraph{Decoupling pseudo-Goldstone modes.} There are a total of seven Goldstone modes. Four are eaten by the massive Standard Model gauge bosons and the dark\,$Z$. We decouple the remaining three by adding a trilinear potential that explicitly breaks the global symmetries of the Higgs sector: 
\begin{align}
  V\supset \mu' H_\text{EW}^\dag H_\text{mix} H_\text{d} + \text{h.c.} \ .
  \label{eq:mu:trilinear}
\end{align}
The scale $\mu'$ is assumed to be large compared to the masses in the three-parameter model. The uneaten Goldstones are an orthogonal linear combination to the eaten combination; see Appendix~A of Ref.~\cite{Chaffey:2019fec}.

\paragraph{Decoupling the dark Higgs}. We may tune the $H_\text{d}$ potential such that that \acro{CP}-even state has the largest mass in the \acro{UV} theory, we define this to be $\Lambda$. When integrating out this mode, the trilinear coupling \eqref{eq:mu:trilinear} generates four-point interactions between the Higgs doublets with a quartic coupling of order $(\mu'/\Lambda)^2$. In a given \acro{UV} model, the perturbativity of the lower-energy two-Higgs doublet theory requires this coupling to be smaller than $\sim 4\pi$.

\paragraph{The decoupling limit of the two-Higgs doublet model.} Upon integrating out $H_d$, our Higgs sector is a \acro{CP}-conserving two-Higgs doublet model. Gunion and Haber examine the viability of a decoupling limit these models in~\cite{Gunion:2002zf}. In their notation, $\Phi_1 = H_\text{EW}$ and $\Phi_2=H_\text{mix}$ and the parameters of their general potential are:
\begin{align}
  m_{12}^2 &= \frac{\mu' v_d}{\sqrt{2}}
  &
  \lambda_{4,5} &=  \frac{\mu'^2}{\Lambda^2}
  & \lambda_{3,6,7} &= 0 \ .
\end{align}
The diagonal quadratic and quartic terms are set to produce symmetry-breaking potentials for each of $\Phi_{1,2}$ that yield the appropriate vacuum expectation values, $v_\text{EW}$ and $v_\text{mix}$, with mixing angle $t_\beta = v_\text{mix}/v_\text{EW} \ll 1$. The mixing angle $\beta$ encodes the same information as our mass mixing parameter, $\kappa$, \eqref{eq:mass:mixing:definition}. This potential is manifestly stable in all field directions. The masses for the \acro{CP}-odd and charged Higgses---the uneaten would-be Goldstones--are 
\begin{align}
  m_{A}^2 = m_{H^\pm}^2 = \frac{\mu' v_d}{\sqrt{2} s_\beta c_\beta} - \frac{v^2  \mu'^2}{\Lambda^2} \gg v^2 \ .
\end{align}
The mass matrix between \acro{CP}-even Higgses is, to leading order in $s_\beta \ll 1$,
\begin{align}
  \mathcal M^2 = 
  \begin{pmatrix}
    \lambda_\text{EW} v^2 & \left(\frac{\mu'^2}{\Lambda^2} -m_{A}^2\right)s_\beta \\
    \left(\frac{\mu'^2}{\Lambda^2} -m_{A}^2\right)s_\beta & m_{A}^2 \\
  \end{pmatrix}
\end{align}
where $\lambda_\text{EW}$ is the quartic coupling for $\Phi_1 = H_\text{EW}$. The rotation to diagonalize the \acro{CP}-even mass matrix is $\alpha$. The lighter eigenstate is identified with the 125~GeV Higgs boson. The decoupling limit corresponds to the case where an entire \acro{SU(2)}$_\text{L}$ doublet is heavy. It corresponds to the limit $(\beta-\alpha) \to \pi/2$, or $m_{A}^2 \gg |\lambda_\text{EW}|v^2$. This is readily realized in our model.

\paragraph{Alternative UV completions.} The minimal model in Refs.~\cite{Gopalakrishna:2008dv, Davoudiasl:2012ag} avoids any extraneous would-be Goldstones. Another alternative is to introduce a Stuckelberg~\cite{Stueckelberg:1938hvi, Feldman:2007wj} mass for the Abelian hidden sector.

\section{Unpolarized Observables}
\label{app:unpolarized:observables}

The angular dependence of the differential $e^+e^-\to A' \to \mu^+\mu^-$ cross section with spin-averaged $e^\pm$ beams is
\begin{align}
    \frac{d\sigma}{d\Omega} &= 
    \frac{f(s)}{4}
    \left[ 
        \left(g_\text{L}^4+ g_\text{R}^4\right)\left(1+\cos\theta\right)^2
        +
        2g_\text{L}^2 g^2_\text{R}\left(1-\cos\theta\right)^2
    \right] \ ,
    \label{eq:unpolarized:diff:xsec:1}
\end{align}
where $f(s)$ is a kinematic factor independent of the scattering angle of the outgoing muon with respect to the electron beam, $\cos\theta$. The forward and backward and cross sections are
\begin{align}
    \sigma_\text{F} &\equiv \int_0^1  d\cos\theta\, \frac{d\sigma}{d\cos\theta}
    &
    \sigma_\text{B} &\equiv \int_{-1}^0 d\cos\theta \frac{d\sigma}{d\cos\theta} 
     \ .
    \label{eq:forward:backward}
\end{align}
The forward--backward asymmetry is expressed in terms of a coupling asymmetry $A_e$,
\begin{align}
    A_\text{FB}^0
    &\equiv
    \frac{ 
        \sigma_\text{F} - \sigma_\text{B} 
        }{ 
        \sigma_\text{F} + \sigma_\text{B} 
        }
    =
    \frac{3}{4}
    A_e^2
    &
    A_e \equiv \frac{g_\text{L}^2 - g_\text{R}^2}{g_\text{L}^2 + g_\text{R}^2}  = 
    \frac{2q_\text{A} q_\text{V}}{q_\text{A}^2 + q_\text{V}^2}
    \label{eq:AFB}
    \ ,
\end{align}
where the superscript on $A_\text{FB}$ indicates an unpolarized asymmetry.

\section{Polarized Beams and Observables}
\label{app:polarized:beams}

We review the collider phenomenology of polarized beams following the conventions in Ref.~\cite{Moortgat-Pick:2005jsx}. Define the longitudinal electron beam polarization to be $P=(R-L)/(R+L)$ where $R$ $(L)$ is the number of right (left) helicity particles in the beam. Similarly, let $\bar P$ be the analogous quantity for the positron beam. The polarized cross section $\sigma_{P\bar P}$ is related to the pure helicity cross sections $\sigma_{\text{R}\bar{\text{L}}}$ and $\sigma_{\text{L}\bar{\text{R}}}$ as
\begin{align}
  \sigma_{P\bar P} = 
  \frac{1}{4}(1+P)(1-\bar P)\sigma_{\text{R}\bar{\text{L}}}
  + 
  \frac{1}{4}(1-P)(1+\bar P)\sigma_{\text{L}\bar{\text{R}}}\ ,
\end{align}
where we used $\sigma_{\text{L}\bar{\text{L}}} = \sigma_{\text{R}\bar{\text{R}}}$ in the limit of massless fermions coupled to a a vector boson. It is convenient to write this in terms of the symmetric (anti-symmetric) combinations 
\begin{align}
\sigma_{(\text{LR})} &= \sigma_{\text{L}\bar{\text{R}}} + \sigma_{\text{R}\bar{\text{L}}} = 4\sigma
&
\sigma_{[\text{LR}]} &= \sigma_{\text{L}\bar{\text{R}}} - \sigma_{\text{R}\bar{\text{L}}}
% &\frac{\sigma_{[LR]}}{\sigma_{(LR)}} &= A_{LR} 
\label{eq:sigma:LR:symmetric:antisymmetric}
\ ,
\end{align}
where we note that the spin-averaged (unpolarized) cross section, $\sigma$, is one fourth of $\sigma_{(\text{LR})}$.\footnote{The unpolarized cross section may be written in terms of helicity cross sections as\begin{align*}
    \frac{d\sigma}{d\Omega}
    = 
    \frac{1}{4}
    \left(
    \left.\frac{d\sigma}{d\Omega}\right|_{\text{RL} \to \text{RL}}
    + \left.\frac{d\sigma}{d\Omega}\right|_{\text{LR}\to \text{LR}}
    + \left.\frac{d\sigma}{d\Omega}\right|_{\text{RL}\to \text{LR}}
    + \left.\frac{d\sigma}{d\Omega}\right|_{\text{LR}\to \text{RL}} 
    \right) \ .
\end{align*}} 
This gives a compact expression for the polarized cross section in terms of an effective luminosity $\mathcal L_\text{eff}$ and an effective polarization $P_\text{eff}$
\begin{align}
    \sigma_{P\bar P} 
    &
    \equiv
    2\sigma \frac{\mathcal L_\text{eff}}{\mathcal L}
    \left[
    1 - P_\text{eff} A_\text{LR}\right] 
    &
    \frac{\mathcal L_\text{eff}}{\mathcal L} 
    &\equiv \frac{1-P\bar P}{2}
    &
    P_\text{eff} &\equiv \frac{P - \bar P}{1-P\bar P}
    \label{eq:polarized:xsec:shorthand}
    \ .
\end{align}
The effective polarization is zero for unpolarized beams and $\pm 1$ for perfectly polarized beams. 

The left--right asymmetry for perfectly polarized beams is $A_\text{LR}^\text{perfect} = \sigma_{[LR]}/\sigma_{(LR)}$. For partially polarized beams, we assume that one beam is mostly left-helicity while the other is mostly-right helicity. The cross section  the mostly-left helicity electron configuration, $\sigma_{-+}$, is related to that of the mostly-right helicity electron configuration, $\sigma_{+-}$, by the replacements $P\to -P$ and $\bar P \to -\bar P$. The left--right asymmetry for the partially polarized beams is
\begin{align}
    A_\text{LR} &= 
    \frac{1}{P_\text{eff}} 
    \frac{\sigma_{[-+]}}{\sigma_{(-+)}}
    =
    \frac{1}{P_\text{eff}} 
    \left(\frac{|P|+|\bar P|}{1+|P \bar P|}\right)
    \frac{\sigma_{[\text{LR}]}}{\sigma_{(\text{LR})}}
    =
    \frac{1}{P_\text{eff}} 
    \left(\frac{|P|+|\bar P|}{1+|P \bar P|}\right)
    A_e
    \ ,
\end{align}
where $\sigma_{(-+)}$  and $\sigma_{[-+]}$ are the symmetric and antisymmetric combinations of the partially polarized cross sections analogous to \eqref{eq:sigma:LR:symmetric:antisymmetric}. These are related to the helicity cross sections by
\begin{align}
     \sigma_{(-+)}
     &= 
     \frac{1}{2}\left(1+|P \bar P|\right)\sigma_{(\text{LR})}
     &
     \sigma_{[-+]}
     &=
    \frac{1}{2} 
    \left(|P|+|\bar P|\right) \sigma_{[\text{LR}]} \ .
\end{align}

One may measure the forward--backward asymmetry for polarized beams. In this case, \eqref{eq:unpolarized:diff:xsec:1} is replaced with
\begin{align}
  \frac{d\sigma}{d\Omega} &=\phantom{+} \frac{f(s)}{4}
  \left[ g_L^4 (1-P)(1+\bar P) + g_R^4(1+P)(1-\bar P)\right](1+\cos\theta)^2
  \\
  &\phantom{=}+ \frac{f(s)}{4} g_L^2g_R^2
  \left[(1+P)(1-\bar P) + (1-P)(1+\bar P)\right](1-\cos\theta)^2 \ .
\end{align}
The resulting forward-backward asymmetry for partially polarized beams is
\begin{align}
  A_\text{FB} = \frac{3}{4} 
  \frac{(g_\text{R}^2 -g_\text{L}^2)^2 + P_\text{eff}(g_\text{R}^4 - g_\text{L}^4)
  }{
  (g_\text{R}^2+g_\text{L}^2)^2 + P_\text{eff}(g_\text{R}^4 - g_\text{L}^4)}
  =
  \frac{3}{4}\frac{(A_e-P_\text{eff})A_e}{1-P_\text{eff}A_e} \ .
\end{align}

%% Bibliography
\bibliographystyle{utcaps} 	% arXiv hyperlinks, preserves caps in title
\bibliography{ILC_DarkPhoton}

\end{document}